\documentclass{aastex63}

\usepackage[english]{babel}
\usepackage[utf8]{inputenc}
\usepackage[T1]{fontenc}

\let\tablenum\undefined

\usepackage{amsmath}
\usepackage{graphicx}
\usepackage[caption=false]{subfig}
\usepackage{enumerate}
\usepackage{siunitx}

\newcommand{\diff}{\mathrm{d}}

\newcommand{\Bjoern}[1] {\textcolor{black}{#1}}
\newcommand{\Antonius}[2]{\textcolor{black}{#2}}

\shorttitle{Interpolation of turbulent magnetic fields}
\shortauthors{Schlegel et al.}

\begin{document}
\title{Interpolation of Turbulent Magnetic Fields and its Consequences on Cosmic Ray Propagation}
\author{L.\ Schlegel}
\author{A.\ Frie}
\author{B.\ Eichmann}
\author{P.\ Reichherzer}
\author{J.~Becker Tjus}
\affiliation{Ruhr Astroparticle and Plasma Physics Center (RAPP Center), Ruhr-Universit\"at 
 Bochum, Institut f\"ur Theoretische Physik IV/ Plasma-Astroteilchenphysik, 44780 Bochum, Germany}

\begin{abstract}
Numerical simulations of the propagation of charged particles through magnetic fields solving the equation of motion often leads to the usage of an interpolation in case of discretely defined magnetic fields, typically given on a homogeneous grid structure. However, the interpolation method influences the magnetic field properties and therefore, also the propagation of particles through these fields. 
In order to determine the resulting error
we compare three different interpolation routines -- trilinear, tricubic and nearest neighbor interpolation -- in the case of isotropic, turbulent magnetic fields. 
First, we analyze the impact of the different interpolation methods on the root mean square field strength, the divergence and the spectrum of the turbulent magnetic field. Here, the nearest neighbor interpolation shows some clear benefits compared to the trilinear method, however, that changes significantly if we consider the particle propagation. In principle, a better interpolation method yields also a better description of the particle transport. 
In the case of field line random walk, it is shown that none of these methods, especially not the nearest neighbor interpolation, is able to yield an accurate description of the diffusion coefficient, exposing the need for a continuous grid-less turbulent magnetic field. 
We optimize the performance of an algorithm that generates such a magnetic field by more than an order of magnitude. Further, the necessary number of wave-modes is determined, so that this continuous method supports realistic simulations over a larger energy range without limitations by the available memory.
\end{abstract} 

\paragraph{\textbf{This is the Accepted Manuscript version of an article accepted for publication in The Astrophysical Journal.  IOP Publishing Ltd is not responsible for any errors or omissions in this version of the manuscript or any version derived from it.  The Version of Record is available online at https://doi.org/10.3847/1538-4357/ab643b}}

\section{Introduction}
Turbulent magnetic fields are present in almost every astrophysical environment and so they yield a key impact on the properties of cosmic rays (CRs)\footnote{CRs are composed of ionized nuclei (predominantly protons) with energies above a few GeV.} that penetrate the Earth's atmosphere. First, these fields play a crucial role at the cosmic accelerators in order to isotropize the particle distribution and afterwards, during the propagation to Earth, turbulent magnetic fields often provide the dominant magnetic field, especially in the large-scale structure of the Universe, e.g.\ in the filaments and sheets. And even in the case of the Galactic magnetic field, the random magnetic field component provides a significant contribution \citep{Beck_2016}. Thus, the identification of the CR sources is strongly coupled to the understanding of the turbulent magnetic fields. But in particular the extragalactic magnetic fields (EGMFs) are hardly constrained by observations. Their filling factors, which provide the fraction of the total volume filled with magnetic fields higher than a certain reference value, vary between different models by several orders of magnitude \citep{AlvesBatista:2017vob}. Another source of uncertainty that is usually not taken into account, results from the inevitable interpolation of a magnetic field that is discretely defined on a grid. Usually the grid provides a homogeneous structure, defined by a certain number $n$ of grid points on each of which the information on the magnetic field is provided, as well as the spacing $d$ between them. 
Obviously, such a grid structure is unable to cover small-scale variations of the field accurately, e.g.\ the impact of single galaxies and galaxy clusters on the total EGMF structure, due to limited memory space. 
To a certain extent this issue can be solved by using a multi-resolution grid \citep{2016JCAP...08..025M}, but the main problem of the interpolation remains. 
So, even in the case of large-scale variations of the vector field, e.g.\ the turbulent magnetic field structure of the voids, filaments and sheets, the interpolation method always provides an interpolation error depending on the chosen method and the grid resolution, in particular. 
In the case of vectorial data on the grid point, either a component-wise interpolation or an interpolation that separates the direction and the magnitude can be performed. Moreover, also a combination of both is possible, e.g. interpolate component-wise at first and change the magnitude to the mean length of the surrounding vectors through normalization afterwards. In principle, all of these procedures either struggle with the conservation of the mean of the magnitudes given on the grid points or the uniqueness of the interpolated direction. 

A commonly used, fast routine is the \emph{trilinear} interpolation, hereafter abbreviated as TL. 
It is defined via the discretization of the to-be interpolated function (that in principle can be either scalar or vectorial as in our case) $f(x,y,z)$
\begin{equation}
f(x,y,z) = \sum_{i,j,k=0}^{1} f(i,j,k)\,x^{i}y^{j}z^{k}\,(1-x)^{i+1}\,(1-y)^{j+1}\,(1-z)^{k+1}
\label{eq:trilinpol}
\end{equation}
in three dimensions in a unit cube. Here, the coordinates $x,y,z$ in the cube volume denote the position where an approximation of the grid data is calculated. In the case of vectorial data, like magnetic fields, typically each component at the coordinates is treated separately referring to the component-wise interpolation approach. 
However, this interpolation method systematically yields a lower magnetic field strength between the grid points than at the grid points. Thus, CRs typically show deflections that are expected to be systematically smaller than expected. But also the direction-magnitude separated TL routine keeps the extreme values at the grid points, like in the case of scalar fields, and generates structural artifacts as illustrated in Fig. \ref{fig:slice_tl}. In addition, the TL does not keep the magnetic field divergence free.

A routine that includes more information on the given data at the grid points is the \emph{tricubic} interpolation, hereafter abbreviated as TC, which is given by \cite{LeMa}:
\begin{equation}
f(x,y,z) = \sum_{i,j,k = 0}^{N} a_{ijk}x^{i}y^{j}z^{k} \,. \label{eq:cubipol}
\end{equation}
There are different possibilities of how to determine the necessary constraints resulting in the coefficients $a_{ijk}$ gaining different properties of the interpolant (see \cite{LeMa}). We focus on the restriction that the derivative at a certain point is given by the difference quotient of the previous and following grid point defined as $\tau(p_{i+1}-p_{i-1})$, leading to the so called Catmull-Rom-Spline-Interpolation with a tension-parameter $\tau$, often and in our case chosen to be $0.5$ \citep{catmullrom}. 
The higher degree of the cubic polynom makes the interpolated field lines smoother and the extreme values do not have to be at the grid points anymore as illustrated in Fig. \ref{fig:slice_tc}.
The divergence is also not conserved by this approach but expected to be smaller than in the case of TL due to the higher order of the polynom. 

It is also possible to interpolate on the grid data  with less information than in the case of TL, like in the case of the 
\emph{nearest neighbor} interpolation, hereafter abbreviated as NN. Hereby, the data from the closest grid point with respect to the current spatial position is chosen without taking any additional grid points into account (see Fig.~\ref{fig:slice_nn}). This routine is the less computational intense. While also not divergence conservative, it still preserves physical constraints as the root-mean-squared field strength, provided at the grid points. Moreover, there are none of the previously discussed issues with respect to treatment of vectorial data.   

With respect to the huge uncertainties of the Galactic and extragalactic magnetic field structure, the error by the interpolation method is most likely negligible. However, it is crucial to quantify these errors, that are in principle avoidable, and elaborate the most efficient routine taking also the computational time into account. When it comes to significant deflections by the magnetic field in the so-called diffusion regime, the particle distribution is commonly used to infer the diffusion coefficient. Especially, for such an analysis it is crucial to understand the impact of interpolated magnetic field structures on the outcome. 

This work is carried out using the publicly available code CRPropa3 \citep{CRPropa3}, which has been extended in this work by the different interpolation routines discussed above, as well as a continuous (grid-less) turbulent magnetic field structure, as introduced in Sect.\ \ref{TD13}. 
This paper is organized as follows: 
In Sect.~\ref{PropChar} the characteristics of turbulent magnetic field structures and diffusive CR propagation are summarized briefly. In addition, we introduce a method to provide a continuous (grid-less) turbulent magnetic field structure, whose performance has been optimized and compared to the grid-based routines. Afterwards, in Sect.~\ref{sec:CompareMagneticField} the impact of different interpolation routines on the magnetic field properties and the transport of CRs are exposed, showing the benefits of the different routines. 

\newpage
\section{Theory}
\subsection{Turbulent magnetic fields}
\label{PropChar}
The question of the generation and maintenance of large-scale magnetic fields in the Universe is still highly debated and unsolved. The existence of intergalactic, turbulent magnetic fields has two evolutionary scenarios: Either, they result from the evolution of primordial fields under the influence of structure formation or, they are generated by the galactic outflow of magnetic fields by winds (e.g.\ \cite{1994RPPh...57..325K, 2001PhR...348..163G, 2008RPPh...71d6901K}). Both the supercluster medium and the large-scale structures of our Universe are hardly constrained by observations. The two different scenarios may cause huge differences with respect to the magnetic field strength \citep{Hackstein2018}. Currently, only upper limits of about a few nG constrains the field strength in the voids \citep{Pshirkov:2016, Planck2016xix}. Stronger constraints can in principle be derived from the observation of gamma-ray induced cascades \citep{neronov2013} - it is not clear, however, how large the influence of the pair instability during the propagation of the electron-positron pairs are not well-quantified at this point \citep{broderick2012}.

In the following, we suppose for simplicity that a uniform isotropic turbulent magnetic field is present within these large-scale structures and consider the propagation of individual CRs through these fields. The field will be characterized as follows:
\begin{enumerate}[(i)]
    \item the root mean squared strength $B_{\rm rms}=\sqrt{\langle B^2(x) \rangle}$; and 
    \item the distribution of magnetic energy $w$, which is usually given by a power-law in Fourier space, i.e.\ 
    \begin{equation}
    w(k) = \frac{B_{\rm rms}^2}{8\pi}\, k^{-m}\, \frac{(m-1)\,k_{\rm min}^{m-1}}{1-(k_{\rm max}/k_{\rm min})^{m-1}}\,,
    \label{eq:wk}
    \end{equation} 
    between a minimum and maximum wave-number, $k_{\rm min}$ and $k_{\rm max}$, respectively. 
\end{enumerate} 
In the following a Kolmogorov spectrum of turbulence with $m=5/3$ will be considered, where the initial energy is injected at a maximum scale $l_{\rm max}=2\pi/k_{\rm min}$. Here, it is transferred by wave interactions to lower scales until it dissipates at $l_{\rm min}=2\pi/k_{\rm max}$. Note, that the spectrum of actual magnetic field data show that at large scales $l>l_{\rm max}$ the turbulence gains energy yielding a spectral index of $m\simeq 1$ referring to the so-called energy range (e.g.\ \cite{2009ASSL..362.....S}). At small scales $l<l_{\rm min}$ the dissipation of energy yields a steep spectrum with $m\simeq 3$ referring to the so-called dissipation range. The characteristic scale on which the magnetic field will vary, the so-called coherence length $l_{\rm c}$ is given in the case of uniform isotropic turbulence by (e.g.\ \cite{2014Harari})
\begin{equation}
    l_{\rm c} = \frac{8\pi^2}{B_{\rm rms}^2} \int_0^\infty \frac{\diff k}{k}\,w(k) = \frac{l_{\rm max}}{2}\,\frac{m-1}{m}\,\frac{1-(l_{\rm min}/l_{\rm max})^{m}}{1-(l_{\rm min}/l_{\rm max})^{m-1}}\simeq \frac{l_{\rm max}}{5}\,,
\end{equation}
where the latter uses $l_{\rm min}\ll l_{\rm max}$ and Kolmogorov turbulence. An efficient interaction with the magnetic wave-modes only happens, if the Larmor radius $R_{\rm L}=E/(Ze\,B_{\rm rms})$ of the CR with an energy $E$ and charge $Ze$ is smaller than the coherence length. Hence, the dimensionless rigidity $\rho=R_{\rm L}/l_{\rm c}$ is the crucial quantity of the CR particle with respect to its interaction with the magnetic field. 

The particles undergo resonant scattering with the turbulent wave-modes for a pitch-angle cosine $\mu$ that satisfies $\mu\,k=1/R_{\rm L}$. 
Thus, $k_{\rm min}\leq 1/R_{\rm L}$ leads to the critical rigidity $\rho=1$, which divides the resonant regime at low rigidities from the quasi-ballistic regime at high rigidities. So, at $\rho\gg 1$ the CRs propagate quasi-rectilinear providing deflections after traversing a distance $l_{\rm c}$ of the order $\sim l_{\rm c}/R_{\rm L}$. Further, the resonant scattering with the turbulent wave modes is only possible in case $k_{\rm max}>1/R_{\rm L}$ leading to the minimal rigidity criterion $\rho\gtrsim l_{\rm min}/l_{\rm max}$. Hence, the impact of diffusive particle transport is the strongest at the \emph{resonant regime} at 
\begin{equation}
    l_{\rm min}/l_{\rm max}\lesssim \rho\lesssim 1\,.
    \label{resonantScattering}
\end{equation}
Here, the characteristic quantity is given by the diffusion coefficient, which determines the spatial distribution of the particles, as introduced in the following section. 
At small rigidities ($\rho\ll 1$) the particle transport is also effected by field line random walk, where the test particle follow the magnetic field lines that diffuse in space (e.g.\ \cite{Subedi_2017, PhysRevLett.21.44}). Note that especially in the case of $\rho\ll l_{\rm min}/l_{\rm max}$, field line random walk becomes the dominant transport process.
\subsection{Diffusive Transport}
In principle, the spatial distribution of charged particles in three-dimensional space after sufficient long propagation time is described by the diffusion equation, neglecting all influences except the isotropic turbulent magnetic field (e.g.\ \cite{Berezinskii1990, Schlickeiser2002, Shalchi2009})
\begin{align}\label{3d diff}
\frac{\partial N(x,y,z,t)}{\partial t}=\kappa_{xx} \frac{\partial^2 N(x,y,z,t)}{\partial x^2} + \kappa_{yy} \frac{\partial^2 N(x,y,z,t)}{\partial y^2} + \kappa_{zz} \frac{\partial^2 N(x,y,z,t)}{\partial z^2},
\end{align}
where the spatial diffusion coefficient is defined as 
\begin{align}
\kappa_{ii} = \frac{v_i^2}{8} \int_{-1}^{+1} \mathrm{d} \mu \frac{(1-\mu^2)^2}{D_{\mu \mu}}.
\end{align}
Here, $\mu$ denotes the cosine of the pitch angle, $v_i$ is the particle velocity which is subsequently estimated by the speed of light $c$, and $D_{\mu\mu}$ is the diffusion coefficient (see \cite{Shalchi2009} for a detailed review).
The mean free path $\lambda = 3\kappa /c$ determines the time required until the diffusive propagation regime is reached and the particle distribution can be described by the diffusion equation in the limit of relativistic particle velocities.  
In the case of an isotropically emitting point source, i.e.\ $N(x,y,z,0) = N_0\delta(x)\delta(y)\delta(z)$, the separation approach $N(x,y,z,t) = \rho(t)P(x)P(y)P(z)$ leads to the particle distribution 
\begin{align}
N(R,t,\kappa) = \frac{N_0}{8 \sqrt{\pi^3\kappa^3 t^3}}\cdot \exp\left(-\frac{R^2}{4\kappa t}\right)
\label{diffDistr}
\end{align}
at time $t$ larger than $\lambda/c$ at a distance $R=\sqrt{x^2+y^2+z^2}$.  
Here, it is used that for isotropic turbulent fields, the diagonal components of the diffusion tensor are all equal, so that $\kappa_{xx} = \kappa_{yy} = \kappa_{zz} = \kappa$. 
By considering the transports, however, in each direction individually (i.e. $y=0$ and $z=0$ for the transport along the $x$-axis) yields 
\begin{align}
    \left\langle (\Delta x_i)^2 \right \rangle =  \int_{-\infty}^{+\infty} \mathrm{d}x_i~x_i^2 N(x_i,t,\kappa_{ii}) = 2t\kappa_{ii}
\end{align}
for the second moment along the $x_i$-axis. Rearranging of this equation yields
\begin{align}
    \kappa_{ii} = \lim_{t \to \infty} \frac{\left\langle (\Delta x_i)^2 \right \rangle}{2t},
    \label{diffTens}
\end{align}
where it is applied that the particle distribution $N(x_i,t,\kappa)$ only follows a Gaussian one as presented in Eq.~(\ref{diffDistr}) for ${t \to \infty}$. 
Hence, in the diffusive regime, i.e.\ $t \to \infty$, the mean difference between the proper particle position $\vec{r}_{\rm p}$ and the interpolated one $\vec{r}_{\rm i}$ in the case of an isotropic particle emission can be expressed by
\begin{align}
    |\vec{r}_{\rm p}-\vec{r}_{\rm i}| &\simeq \sqrt{\Delta x^2_{\rm p}-\Delta x^2_{\rm i}+\Delta y^2_{\rm p}-\Delta y^2_{\rm i}+ \Delta z^2_{\rm p}-\Delta z^2_{\rm i}}\\
    &=\sqrt{6(\kappa_{\rm p}-\kappa_{\rm i})/c}\,D_{\rm traj}^{1/2}\,.
    \label{diff_posDiff}
\end{align}
Here, we used the Eq.~\ref{diffTens} for the case of isotropic turbulence and expressed the time by the particles' trajectory length $D_{\rm traj}$.

\newpage
\section{Method}
\label{Sec:Method}
\subsection{Turbulent magnetic fields without interpolation}
\label{TD13}
\subsubsection{Procedure}
In general, interpolation appears to be unavoidable when running simulations on grid-based magnetic fields. However, when the \Bjoern{field} is derived from a mathematical procedure, the grid is not necessarily needed as the procedure itself provides the field at any point in space. 

According to Sect.\ \ref{PropChar}, Kolmogorov-type turbulence is defined by a power-law distribution of wave-modes in the so-called k-space. Technically speaking, the grid-based turbulent magnetic fields are usually generated with the help of an Inverse Fast Fourier Transform (IFFT): A grid in k-space is populated with values according to the given power-law behavior (\ref{eq:wk}); then, an IFFT is used to obtain the corresponding grid in x-space, which provides the turbulent magnetic field.

However, in a method pioneered by \cite{gj99} and improved on by \cite{td13}, hereafter referred to as TD13, the grid is eliminated completely. During setup, the wave-modes are generated in a way similar to the grid-based method, except that (a) the wave-modes are not confined to any grid, and (b) in most cases, significantly \Antonius{less}{fewer} wave-modes are used. At this stage, no field generation in x-space actually occurs.

Instead, when the value of the field is requested at a particular location in space, all of the pre-generated wave-modes are evaluated at that point in space, then added together. This procedure would be mathematically equivalent  to the action of an IFFT when evaluated using grids in k- and x-space, ignoring constant factors\footnote{Though in practice, with this method, grids are neither used in k- nor x-space. Evaluating the typically large number of wave-modes of a k-space grid --- similar to an IFFT --- would not be tractable in terms of run-time: Consider that the IFFT only needs to be performed once, at setup, while this method must run each time the magnetic field is queried.}.

As part of this work, the TD13 method has been implemented in CRPropa following the description in \cite{td13}, except for two differences, as follows:

First, the current implementation uses a straight power-law in k-space, instead of the broken power-law suggested by \cite{td13}. As described there, we could also get rid of the normalization factor due to the built-in normalization procedure. Lastly, since the $k$ are logarithmically spaced, $\Delta k \propto k$. The $\Delta k$ is part of the normalization, so the constant factor can be ignored. Thus, our implementation uses

\begin{equation}
    A^2(k_n) = G(k_n) k_n \left(\sum_{\nu = 1}^{n_m} G(k_\nu) k_\nu\right)
\end{equation}
where $n_m$ denotes the number of wave-modes and $G(k) = k^{-m}$ with the spectral index $m$ of the turbulence spectrum. 
Note that this function yields a constant spectral behavior which is subsequently used to describe the inertial range of the spectrum. In the case of a more realistic turbulence spectrum which also covers the energy and the dissipation range, respectively, the form of $G(k)$ needs to be adopted as given by \cite{2009AdSpR..43.1429S}. However, the artificial case of a steep cut off beyond the inertial range of the spectrum enables to carve out the impact by the resonant scattering regime.

Second, TD13 normalize the field to $B_\text{rms} = 1$ (unitless). To obtain a field that reproduces a certain $B_\text{rms}$, we simply multiply the desired $B_\text{rms}$ onto the normalized formula. 

In total, the continuous isotropic turbulent magnetic field in x-space is given by

\begin{equation}
    \vec{B}(\vec{x}) = \sqrt{2} B_\text{rms}  \sum_{n=1}^{n_m} \vec{\xi_n} A(k_n) \cos (\vec{k_n}  \cdot \vec x + \beta_n)
    \label{TD13_Bx}
\end{equation}

where $\vec{k_n} = k_n \vec{\kappa_n}$, $\beta_n = \zeta_n$; and $\vec{\kappa_n}$, $\vec{\xi_n}$, and $\zeta_n$ are generated according to the process described by TD13.

A visual representation of the resulting field is shown in Fig.\ \ref{fig:slice_td13}.

\subsubsection{Optimization}
\label{sec:optimization}

Since the time needed to evaluate eq. \ref{TD13_Bx}  roughly scales linear with the number of wave-modes, physical accuracy -- requiring more wave-modes -- and optimization of runtime -- requiring fewer wave-modes -- can be traded off against each other. However, an important consideration is the constant factor in this relation, in other words, the amount of time needed per wave-mode. Lowering this amount of time would ease the trade-off described above, by allowing more wave-modes to be computed in less time. Often, the minimal number of wave-modes is set by physical requirements; the corresponding runtime then decides the viability of the simulation.

So we developed an optimized implementation of the previously introduced TD13 turbulent magnetic field due to three technical adjustments: 
\begin{itemize}
    \item We rewrote the implementation to use ``single instruction multiple data'' (or SIMD) processor instructions, which allow each processor core to process multiple numbers in parallel. 
    \item We changed the way the wave-mode data is stored: The code now does some pre-computations at setup time and prepares the data such that each wave-mode can be evaluated with minimal cost.
    \item Lastly, building on the previous step, we sped up the evaluation of the cosine function itself. In particular, we simplified the argument reduction process \citep[see also][]{elementary-functions} by computing $\cos (\pi a)$ instead of $\cos(a)$; we also integrated a custom cosine implementation based on the highly optimized code from \cite{sleef}. 
\end{itemize}

Since the SIMD instructions are not part of the base x86/64 instruction set, and their availability varies across different processors, we provide three different implementations:

\begin{itemize}

\item The non-optimized reference version, which is a straightforward implementation of the formula and is used to verify the optimized versions.

\item The main optimization result, which uses AVX instructions for SIMD and optionally supports the FMA extension for a small performance boost. This configuration should be present on most modern systems.

\item A backport of the AVX version to its predecessor, SSE. Since SSE only allows for simultaneous manipulation of two double-precision floats (as opposed to four in the case of AVX), this version incurs about a 2x performance penalty compared to the AVX version. However, it is still a lot faster than the reference implementation, so this is recommended for older systems which do not support AVX.

\end{itemize}

As an additional note, the AVX version should be treated with care, since it might cause the processor to slightly reduce its clock frequency, which would affect the rest of the simulation as well. In our simulations, this effect was not noticeable for large numbers of wave-modes. However, this may change if more code, such as an interaction module, runs between consecutive queries of the magnetic field. If this slowed the overall simulation down by more than a factor of two, the SSE version would become preferable, since SSE does not affect the CPU clock. Overall, this is something that should only become problematic in a few cases, especially since the AVX frequency reduction can be triggered unintentionally by other parts of the code. In these cases, a quick performance comparison should help select the optimal implementation.
\subsection{Sampling on the grid}
To expose the impact of the different interpolation methods we store each of those previously introduced TD13 field structures on a homogeneous grid with a spacing $d$ and keep the grid volume $(Nd)^3$ constant --- though periodically repeated if needed. 
In doing so, we ensure that the main characteristics of the turbulent magnetic field in the x-space \Antonius{has}{have} not changed. 
In addition, the available memory (we used roughly 8\,GB for the grid storage) and the sampling theorem, which implicates that
\begin{equation}
    l_{\rm min} \geq 2\, d\qquad\text{and }\quad  l_{\rm max} \leq N\, d/2\,,
\label{eq:sampTheo}
\end{equation}
give further constraints on the realization of different sampling resolutions as well as the ratio of $l_{\rm min}/l_{\rm max}$. So, we realize six different cases --- the limiting cases as well as two intermediate ones --- using three different sampling resolutions and three different $l_{\rm min}/l_{\rm max}$ ratios. 
Higher grid resolutions are much more computationally \Bjoern{expensive} and are not likely to provide any additional benefit in the present
study.
\subsection{Analyzing the magnetic field properties}
In order to compare the accuracy of the different interpolation methods we first analyze the resulting properties of the isotropic turbulent magnetic field.

First, we compare the root mean squared field strength $B_{\rm rms}$ that is obtained from the magnetic field strength $B(x)$ at $10^{7}$ arbitrary spatial positions dependent on the interpolation routine with reference to the exact value of $B_{\rm rms} = 1\,\text{nG}$ initially given to create the field. We split the data set into ten parts in order to estimate the statistical uncertainties. Hereby, $B_{\rm rms}$ is calculated for the different routines and field configurations individually. 

Secondly, we \Antonius{like}{would like} to obtain a measure of the B-field divergence in each method. Straight-forward differentiation, however, is not possible: TL is not differentiable, and NN is not even continuous. For this reason, we instead use the Gauss Theorem and compute the net flux through \Bjoern{the surface of small test} regions of space\Antonius{, defined by the grid cells.}{.}

The net flux through any of these test regions \Bjoern{needs to vanish} i.e.\ $\mathrm{div} B = 0$, so that any additional or missing flux represents a flux defect. Since each interpolation method actually defines a representable function (dependent on the grid values, of course, but still a function), the surface flux integration can be done analytically. \Antonius{And since the method is analytical, TD13 is guaranteed to have no divergence (by construction), so it was not evaluated.}{And since TD13 is analytical and by construction free of divergence, it was not evaluated here.}

\Antonius{}{To simplify the computation of the flux defect for the interpolated methods, the test regions are given by the whole grid cells, i.e.\ regions of space enclosed by 8 neighbouring grid points. In particular, this means that the discontinuities produced by NN are included in the computation, since those are located halfway between consecutive grid points.}

\begin{figure}
    \centering
    \includegraphics[width=0.6\linewidth]{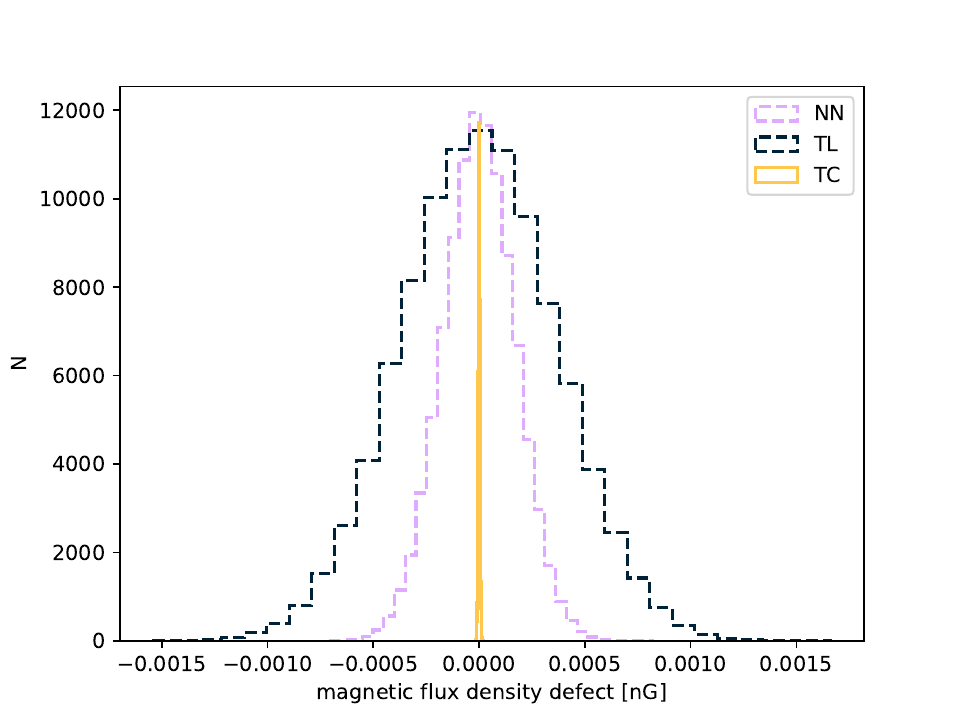}
    \caption{Flux defect density distribution for $10^6$ randomly drawn grid cubes using $l_{\rm min}/l_{\rm max}=0.0972$ and $l_{\rm min}/d=14.4$.} 
    \label{fig:fluxdist-6}
\end{figure}
%

Since the actual \Antonius{}{(physical)} net flux through each test region depends on its surface \Antonius{size}{area}---which can be varied at will, \Antonius{}{by uniformly scaling grid spacing and size}---it does not represent a good measure for error by the interpolation. Hence, we divide the net flux by the surface area to obtain the density of the flux defect. \Antonius{Hence, }{As a result, }this measure becomes independent of the grid size and only depends on intrinsic properties of the underlying system. As the distributions are all centered around zero (see Fig.\ \ref{fig:fluxdist-6}), using the standard deviation $\sigma_B$ of the flux distribution from $10^6$ randomly drawn samples yields the most sensible single-parameter measure of this \Antonius{quantities}{quantity}. 

%

In addition to the interpolation effects in x-space, also the underlying wave spectrum is changed by the interpolation. To analyze the wave spectrum for a given volume of space, \Bjoern{we first sample the continuous TD13 field with a certain resolution on a grid. This grid is subsequently used to determine the magnetic field on some equidistant points in space that are located in the middle of the cubes defined by grid points and we transform these magnetic field vectors into $k$-space by a fast Fourier transform (FFT). Then we create a histogram of the norm of the resulting FFT vectors dependent on their wavenumber $k$ which finally yields the wave spectrum. 
}

\newpage
\subsection{Analyzing the impact on test particle propagation}
In addition to the interpolated magnetic field properties, we also analyze its impact on the transport of a test particle in different propagation regimes. 
First, we compare single-particle trajectories, but to obtain the absolute error of the different methods, the given turbulent magnetic field needs to be used at every point in space to determine the flawless trajectories. Therefore, we use the previously introduced grid-less turbulent magnetic field structure, TD13, based on a given number of wave-modes. 
To provide a single-particle comparison, we propagate 5000 particles with a given rigidity and an isotropic initial momentum distribution through 20 realizations of the turbulent magnetic field and the mean difference that results from these $20\times 5000$ trajectories dependent on the trajectory length $D_{\rm traj}$. Comparing the trajectory $\vec{r}_{\rm TD13}$ that results from the grid-less TD13 field, with the trajectory $\vec{r}$ from the grad based field, where an interpolation routine has been applied, we obtain the relative error of the trajectory
\begin{equation}
\text{err}(\vec{r}) = \frac{|\vec{r}_{\rm TD13}-\vec{r}|}{\vec{D}_{\rm traj}}\,.
\label{relErrorTraj}
\end{equation}
It needs to be taken into account, that the pure impact of the interpolation error only unveils after the very first propagation step. In the following propagation steps the particles' trajectories are also \Antonius{effected}{affected} by the spatial differences of the turbulent magnetic field, which amplifies the difference from the proper particle trajectory. 
In the diffusive regime ($c\,t\gg \lambda$) small deviations by the interpolation get significantly amplified and randomized by the diffusive motion of the particle, so that on average the deviations between the single-particle trajectories is expected to become dominated by the diffusive particle transport rather than by the interpolation imprint.

\begin{figure}[htbp]
\centering
\includegraphics[width=0.32\linewidth]{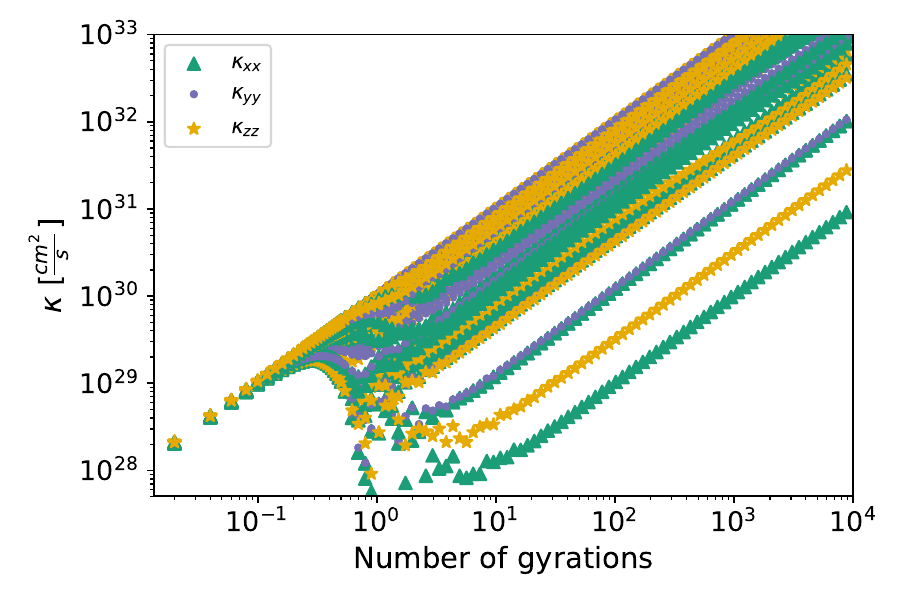}
\includegraphics[width=0.32\linewidth]{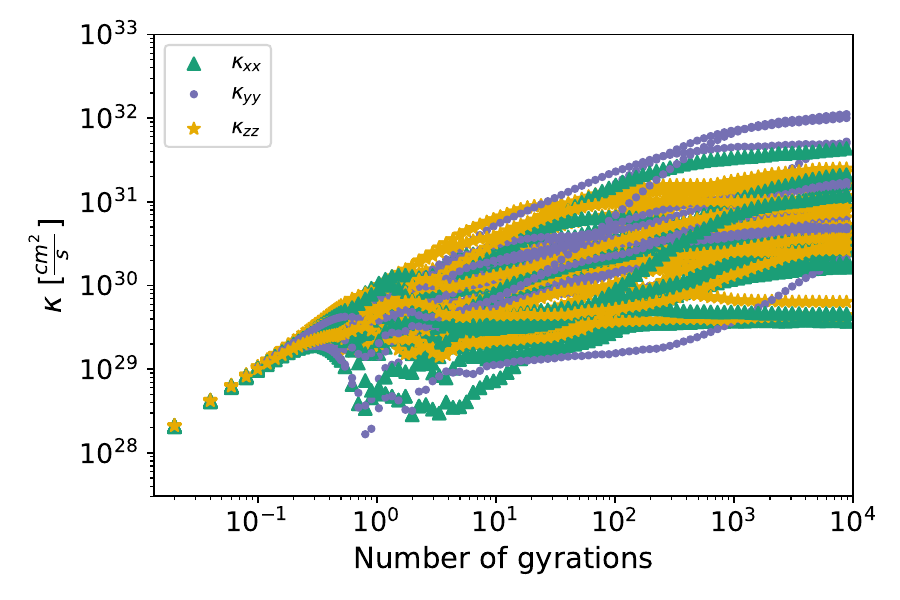}
\includegraphics[width=0.32\linewidth]{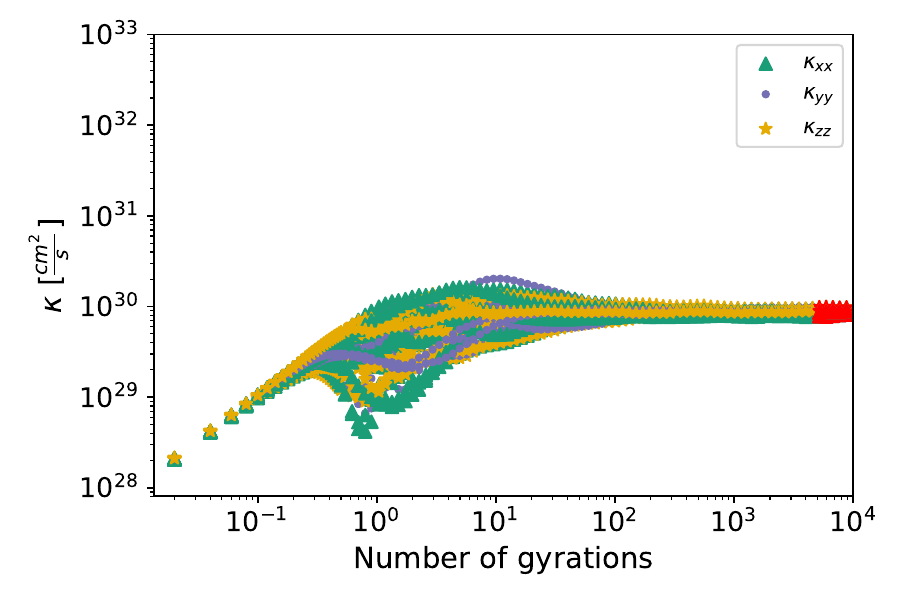}
\caption{Diffusion coefficients $\kappa_{xx},\kappa_{yy}, \kappa_{zz}$ dependent on the numbers of gyrations of a test particle with $\rho=0.1$. Here, the TD13 field is used with 2 (\emph{left}), 5 (\emph{middle}) and 1200 (\emph{right}) wave-modes for 20 random seeds for each plot. Note, that the proper diffusion coefficient is only given at a large number of gyrations, where it has become constant, hence there is no proper diffusion coefficient given in the case of 2 wave-modes (left plot). The red marked area in the right plot marks the last 5\% of the data, which is used to determine the mean diffusion coefficient $\kappa$.}
\label{kappas}
\end{figure}

Finally, we quantify the impact of the interpolation method by its influence on the diffusion coefficient $\kappa$ in the diffusive regime of resonant scattering. Similar to the previous study, we use 5000 isotropically emitted test particles that propagate through 20 different realizations of the same turbulent magnetic field and determine the mean distance $\left\langle (\Delta x_i)^2 \right \rangle$ of these particles along the $x_i$-axis for each of the 20 field realizations. So, the diffusion coefficients $\kappa_{xx}$, $\kappa_{yy}$, $\kappa_{zz}$ are given according to Eq.~\ref{diffTens} for each of the three spatial directions. 
Hereby, we ensure that $c\,t\gg \lambda$, so that the test particle has entered the diffusive regime, where $\kappa$ has reached its plateau and is independent of $t$ (see Fig.~\ref{kappas}). Hence, only the last 5\% of the mean distance data are used in the following to determine the diffusion coefficient. As shown in Fig.~\ref{kappas}, the reference value of the diffusion coefficient which is given by the grid-less turbulent magnetic field structure, TD13, also depends on the chosen number of wave-modes, so that we first need to ensure the use of a sufficient number of wave-modes. Note, that in the case of $n_m<5$ there is no plateau reached after $10^4$ gyrations, so that we cannot determine the proper diffusion coefficient. 
  In the given case of isotropic turbulence $\kappa_{xx}$, $\kappa_{yy}$ and $\kappa_{zz}$ need to become equal in statistical terms, so that we can use the mean diffusion coefficient $\kappa=(\kappa_{xx}+\kappa_{yy}+\kappa_{zz})/3$ in the following. 
Thus, the relative error of the diffusion coefficient is given by
\begin{equation}
\text{err}(\kappa) = \frac{|\kappa_{\rm TD13}-\kappa|}{\kappa_{\rm TD13}}\,,
\label{relErrorDiffCoeff}
\end{equation}
where $\kappa_{\rm TD13}$ denotes the diffusion coefficient that results from the grid-less TD13 field with an appropriate number of wave-modes, and $\kappa$ is the one that yields the grid-based field, where an interpolation routine has been applied.

\section{Results}
In the following, we use an isotropic turbulent magnetic field setup with $B_{\rm rms}=1\,\text{nG}$ and $l_{\rm c}\simeq1\,\text{Mpc}$, as one might expect from the extragalactic, large scale structures in our Universe. Hereby, we used the length scale ratios $l_{\rm min}/l_{\rm max}=\{0.004,\,0.017,\,0.097\}$ of the turbulence spectrum, which enables a grid resolution $l_{\rm min}/d=\{2,\,4,\,14\}$ for $l_{\rm min}/l_{\rm max}=0.097$, $l_{\rm min}/d=\{2,\,4\}$ for $l_{\rm min}/l_{\rm max}=0.017$, and $l_{\rm min}/d=2$ for $l_{\rm min}/l_{\rm max}=0.004$ based on the available memory.
First, we determine the necessary number of wave-modes of the grid-less TD13 field in order to obtain a certain precision and compare the performance of the different routines. 
Afterwards, we expose the impact of the different interpolation methods on the magnetic field properties, as well as on the propagation of particles in the ballistic and diffusive regime.
\subsection{Necessary number of wave-modes}
\label{sec:numOfwave-modes}
\begin{figure}[htbp]
\centering
\includegraphics[width=0.49\linewidth]{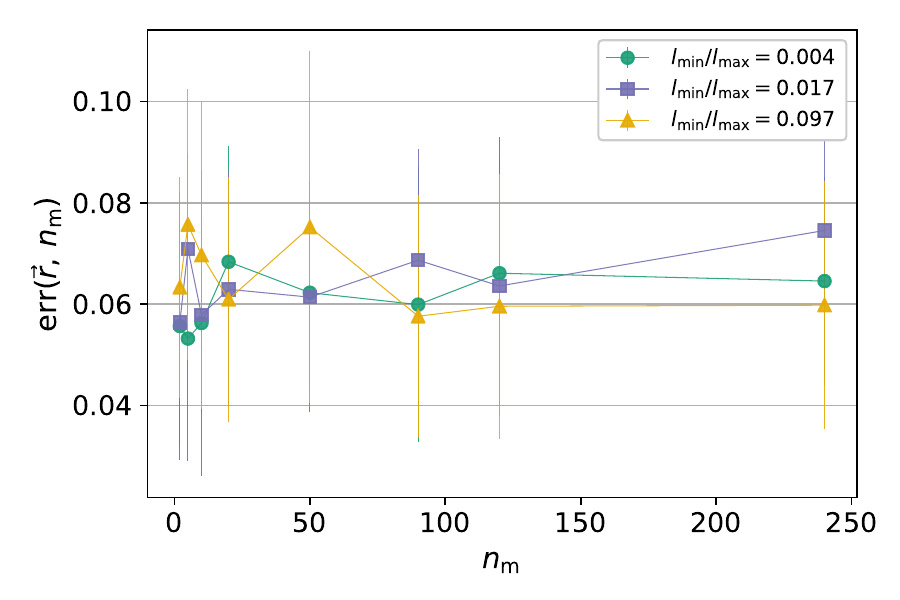}
\includegraphics[width=0.49\linewidth]{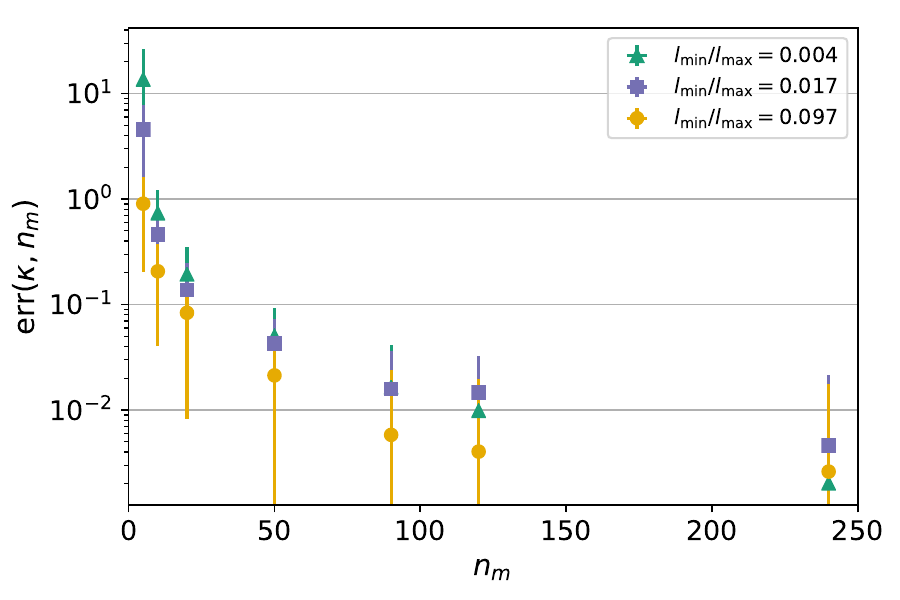} 
\caption{The relative error of the analytical, turbulent magnetic field (TD13) dependent on the number of wave-modes in the case of $\rho=0.1$. \textbf{Left:} Relative difference of the particles' position after the first propagation step, hence within the ballistic regime. \textbf{Right:} The relative error of $\kappa$ in the diffusive propagation regime.}
\label{wave-modes_err}
\end{figure}

In principle, the previously given algorithm (\ref{TD13_Bx}) allows an exact calculation of the magnetic field at any position in x-space only in the case of $n_m\rightarrow \infty$. However, a huge number of wave-modes makes the algorithm computationally intense and therefore difficult to use for CR propagation. Hence, we determine the sufficient number of wave-modes in order to obtain a certain precision. A one to one comparison of the resulting turbulent magnetic fields for different $n_m$ is only possible with respect to their influence on CR propagation. 
Using 5000 particles with a fixed rigidity $\rho=0.1$ that are isotropically emitted and propagate through 20 different realizations of the turbulent magnetic field, we determine the uncertainty
\begin{align}
\text{err}(\vec{r},\,n_m) &= \frac{|\vec{r}_{1200}-\vec{r}_{n_m}|}{\vec{D}_{\rm traj}}\,,\\
\text{err}(\kappa,\,n_m) &= \frac{|\kappa_{1200}-\kappa_{n_m}|}{\kappa_{1200}}
\end{align}
dependent on the number $n_m$ of wave-modes for $l_{\rm min}/l_{\rm max}=\{0.004,\,0.017,\,0.097\}$. Thus, we evaluate the necessary number of wave-modes with respect to the ballistic and the diffusive regime, respectively, where the case of $n_m=1200$ provides the reference values. \newline
In the ballistic regime, we compare the spatial particle position $\vec{r}$ at the very beginning at about $D_{\rm traj} = 14\,\text{kpc}$, where $D_{\rm traj}\ll R_{\rm L}$. It is clearly shown in the left Fig.~\ref{wave-modes_err}, that within the ballistic regime, the initial error is at a few percentage level independent of the number wave-modes. Hence, there is no resonant interaction between the CR and the wave-modes, which would imply such a correlation. However, in the diffusive regime which the particle enters after a multitude of gyro motions ($c\,t\gg R_{\rm L}$), the relative error of the diffusion coefficient depends significantly on $n_m$ as shown in the right Fig.~\ref{wave-modes_err}. Here, the error decreases with an increasing effective number $\tilde{n}_m = n_m/(l_{\rm max}/l_{\rm min})$ of wave-modes, hence, the particles interact with the given number of wave-modes to perform a diffusive motion. As shown in Fig.~\ref{kappas}, the beginning of the diffusive particle motion is shifted to later times in the case of a small number of wave-modes. Using only two wave-modes, the particle is even after $10^4$ gyrations not scattered off these modes, so that the diffusion coefficient cannot be determined. But with a effective wave-mode number of $\tilde{n}_m\geq 1$ we already keep the relative error of the diffusion coefficient below about 1\%, hence it is suggested to keep the used number of wave-modes larger than the length scale ratio of the turbulence. 
Changing the particle's rigidity changes the temporal onset of the diffusive particle motion, but not the necessary number of wave-modes.

\subsection{Performance}\label{sec:perf}

To measure the performance of each method in the overall system, a simple test simulation was set up. The simulation would propagate a certain number of particles through the magnetic field using the \verb/PropagationCK/ module. To exclude unwanted side effects and achieve comparability between methods, each particle was propagated for the exact same number of simulation steps; this was achieved by setting the step size to a constant value and imposing a maximum trajectory length as a stop condition. The CPU time for each execution run was measured with Linux's \verb/time/ utility in a ``best-of-3'' system: Each configuration was run three times; at the end, the fastest value was taken. All simulations were performed on the same computer setup.

\begin{figure}[htbp]
    \centering
    \includegraphics[width=0.8\linewidth]{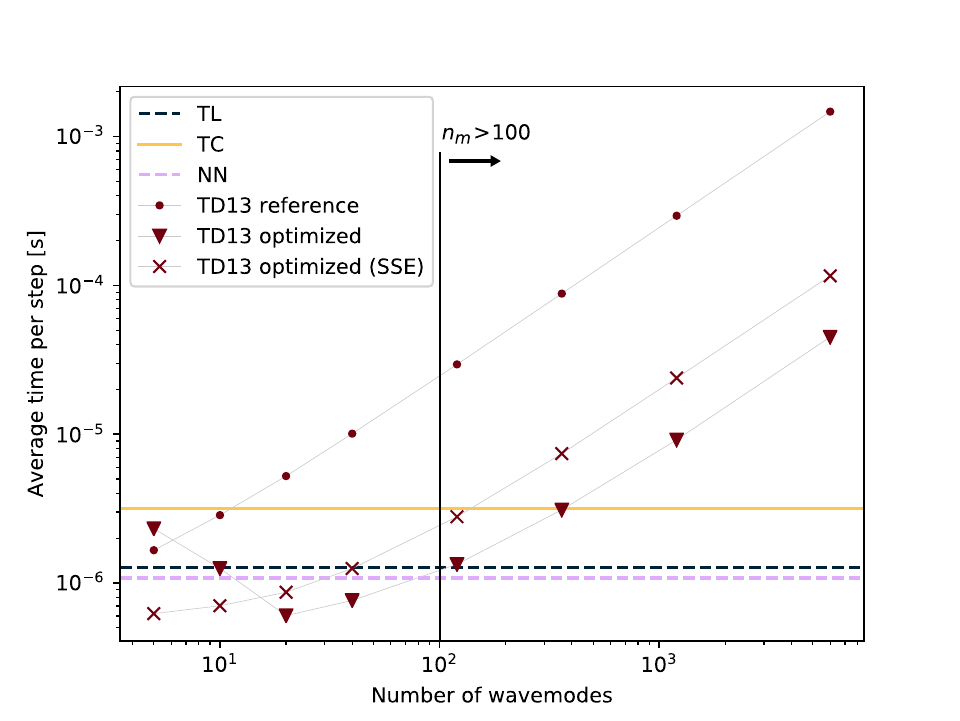}
    \caption{Performance of various magnetic field methods in test simulations. The grid-based methods (TL, TC, NN) do not have a parameter that affects their runtime, so they are drawn as horizontal lines. Data for TD13 is shown for different numbers of wave-modes and our recommendation for $n_m > 100$ is indicated by the vertical line.
    \label{fig:perf} 
    }
\end{figure}

In order to avoid unwanted influence from setup-time operations (such as generating the grid for grid-based methods), each method was run with \num{1e4}, \num{2e4}, \num{3e4}, and \num{4e4} particles. Using linear regression, the slope of a fit through these points was obtained, yielding the average time per particle. This was then divided by the number of steps for each particle to obtain the average time per step, shown in Fig.\ \ref{fig:perf} as a function of the number of wave-modes.

TD13 was tested on multiple counts of wave-modes, since these influence performance; the grid-based methods do not depend on a parameter that affect their runtime, so they are shown as horizontal lines. 

With respect to the different interpolation routines the computational time increases with increasing interpolation effort, so that TC takes about three times longer than NN. The grid-less TD13 field shows the longest computation time in the case of the non-optimized algorithm for $n_m>10$. In the case of the recommended number of wave-modes of $n_m>100$, the routine is already ten times slower than TC, hence hardly feasible for the propagation of a sufficient number of particles. However, with the previously introduced optimization, the computation time can be reduced by more than an order of magnitude for $n_m\gtrsim 20$. Thus, the performance of the most-optimized TD13 algorithm for 100 wave-modes equals the one of TL, so that there is in principle no benefit in restricting the simulations to grid-based turbulence fields.

\subsection{Magnetic field properties}
\label{sec:CompareMagneticField}
\begin{figure}
    \centering
    \includegraphics[width=1.0\textwidth]{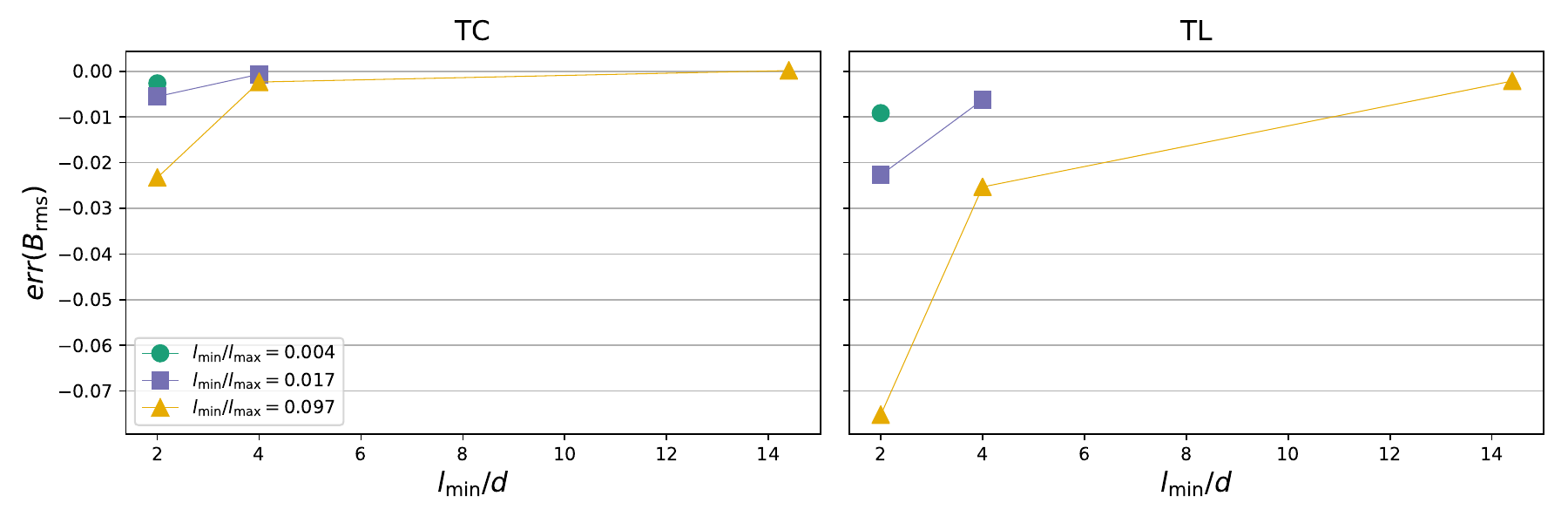}
    \caption{Relative deviation of the $B_{\rm rms}$ in comparison to the initially given 1\,nG for different $l_{\rm min}/l_{\rm max}$ ratios and samplings (as far as the setup allowed it to have data points) for the interpolations TL and TC, using $10^{7}$ randomly drawn values each plotted data point.}
    \label{fig:brms}
\end{figure}

\begin{figure}
    \centering
    \includegraphics[width=\linewidth]{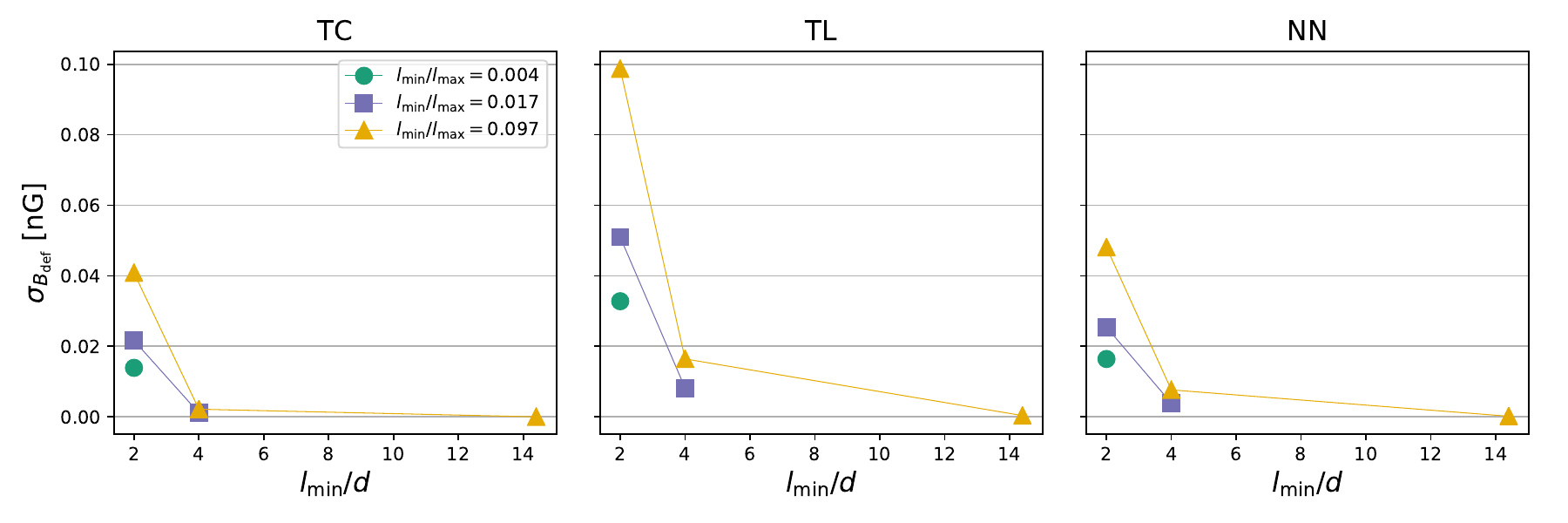}
    \caption{The standard deviation of the flux defect density distribution dependent on the field configuration as well as the interpolation method. A lower value of $\sigma_{B_{\rm def}}$ corresponds to a more divergence-free magnetic field overall.}
    \label{fig:fluxdist-staircase}
\end{figure}

In the following, we compare different properties, like the root-mean-squared field strength and the magnetic flux defect, of the isotropic, turbulent magnetic field that results from the different interpolation routines.
\begin{figure}[hpbt]
\centering
\subfloat[TD13 method]{
\includegraphics[width=0.45\textwidth]{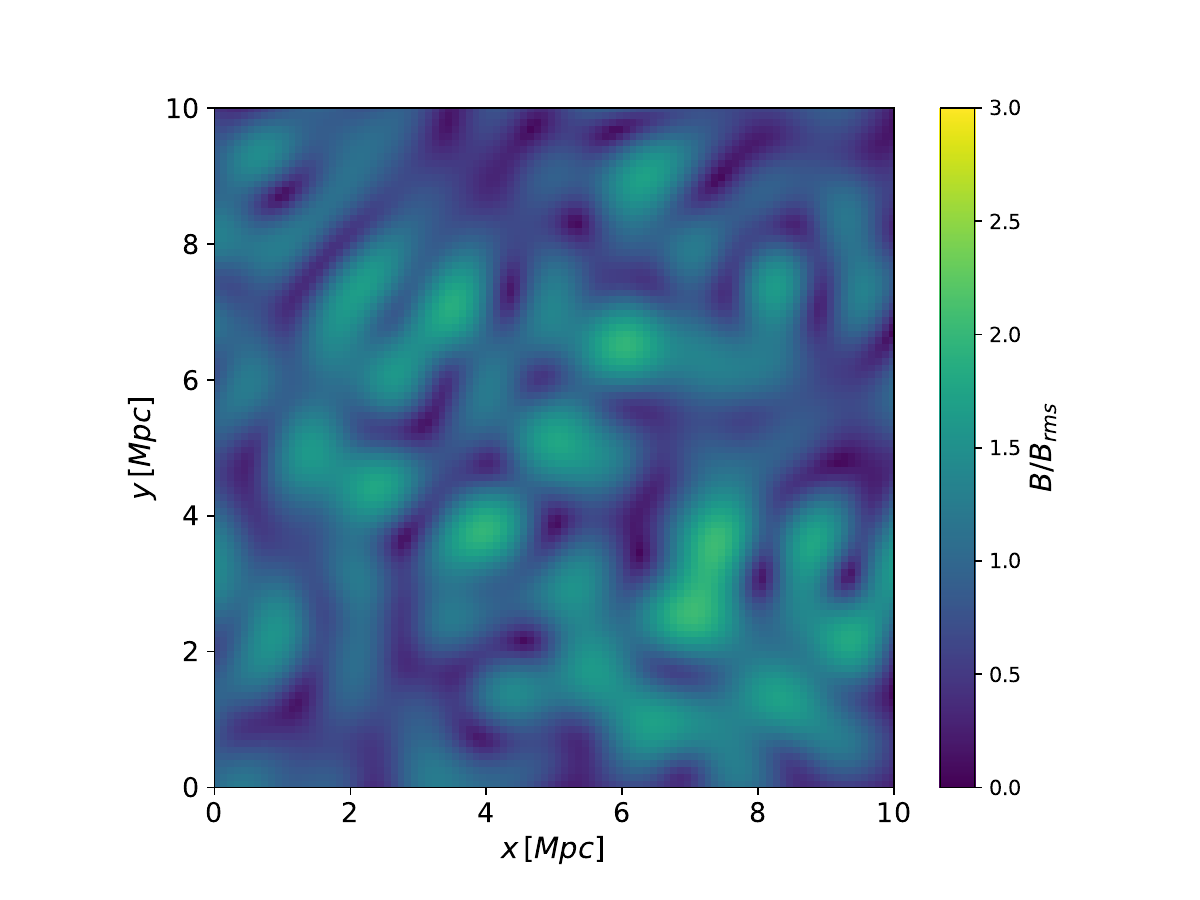}
\label{fig:slice_td13}
}
\subfloat[Tricubic interpolated]{
\includegraphics[width=0.45\textwidth]{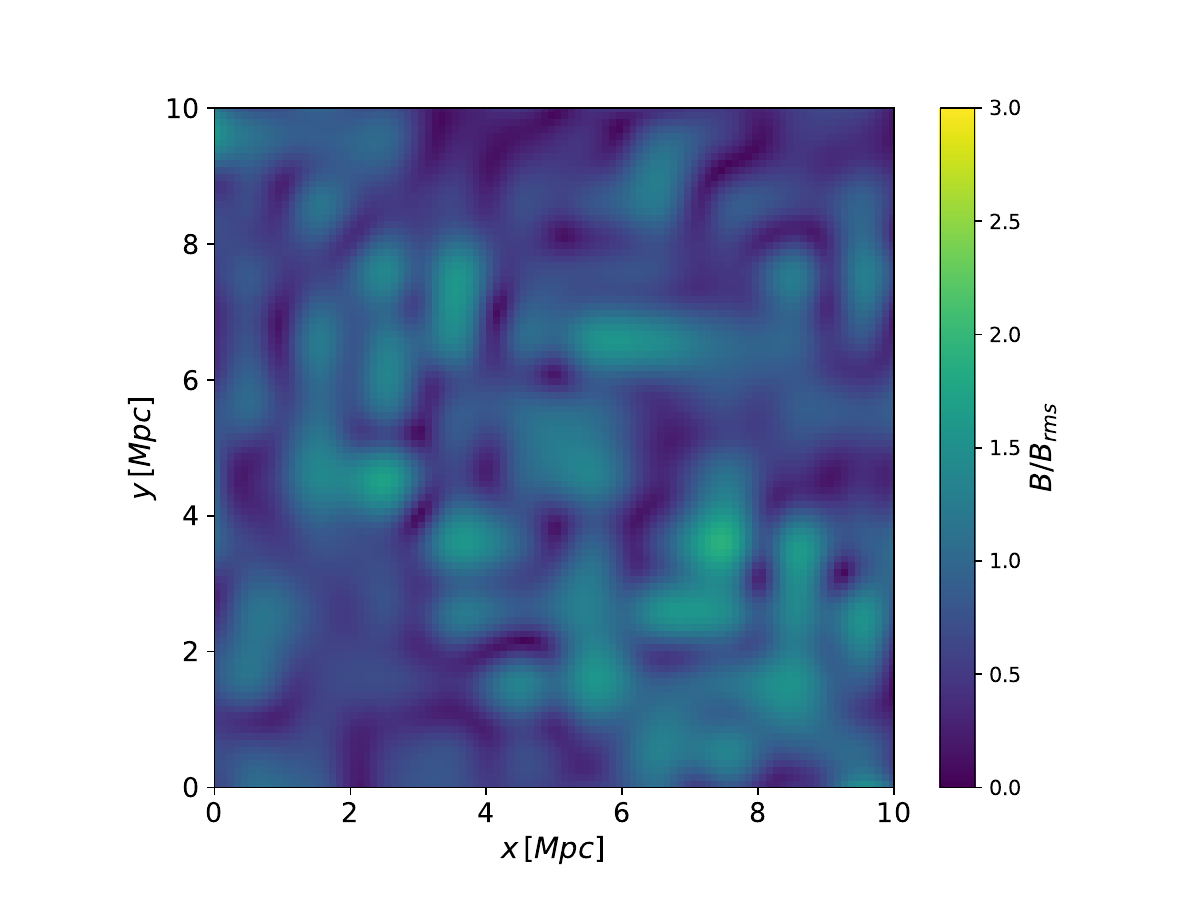}
\label{fig:slice_tc}
}\\
\subfloat[Trilinear interpolated]{
\includegraphics[width=0.45\textwidth]{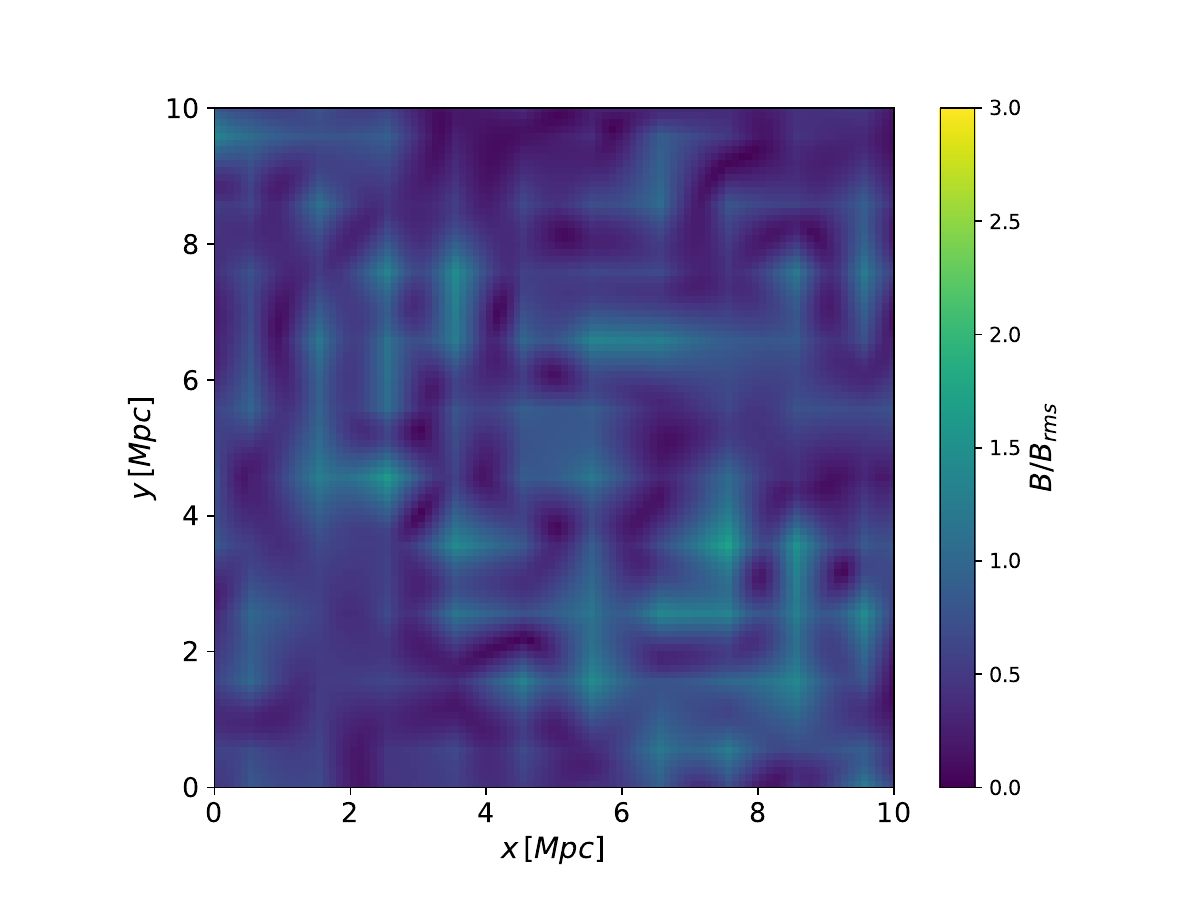}
\label{fig:slice_tl}
}
\subfloat[Nearest neighbor interpolated]{
\includegraphics[width=0.45\textwidth]{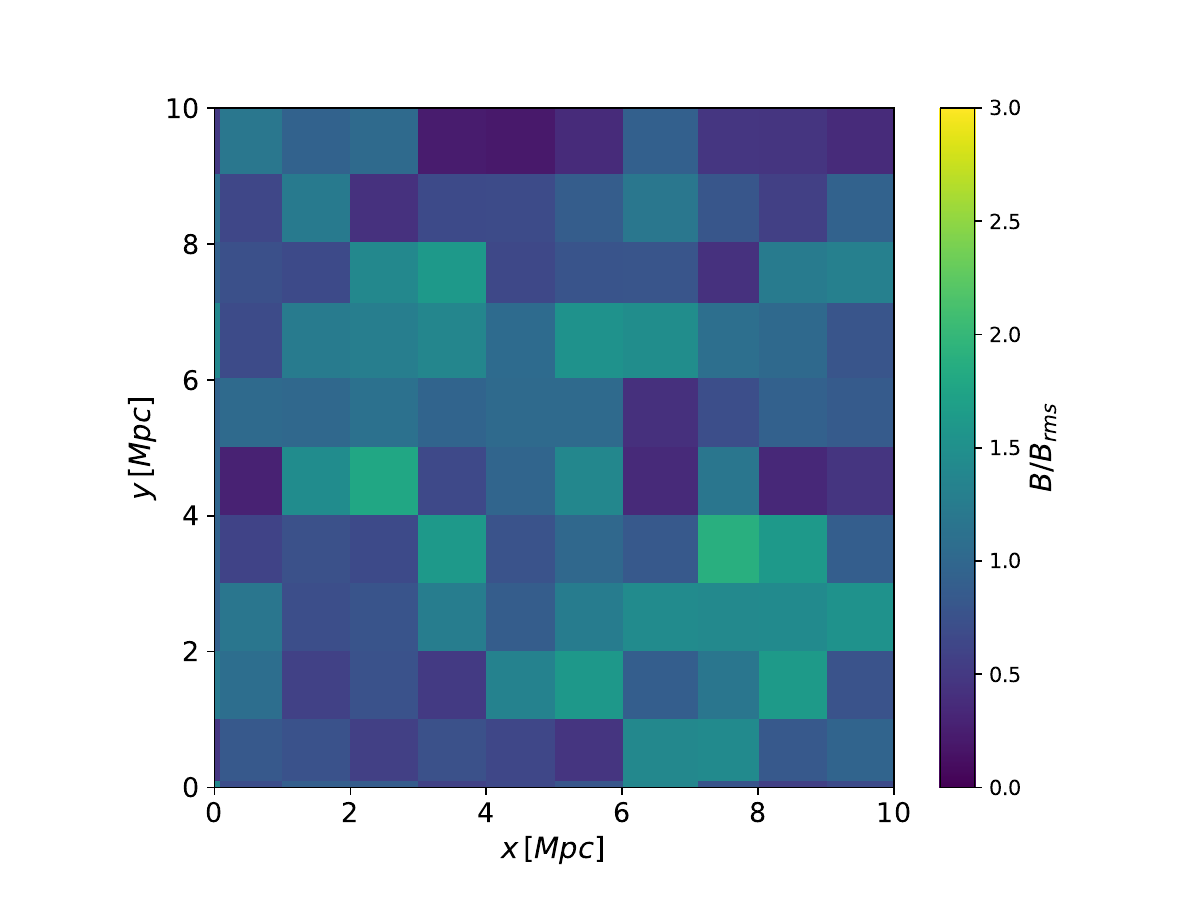}
\label{fig:slice_nn}
}
\caption{Sliceplots (at z=64\,Mpc) showing an extract of the same turbulent field with $\lambda=1\,$Mpc generated with TD13, sampled and interpolated with the three different methods TC,TL,NN. The color represents the magnetic field strength normalized by the root-mean-squared strength.}
\label{fig:sliceplots}
\end{figure}
Figure \ref{fig:sliceplots} shows a cut-out of a certain realization of the magnetic field structure given by TD13 (Fig. \ref{fig:sliceplots}a), as well as the resulting magnetic field structure from the different interpolation methods (a: TD13; b: TC, c: TL and d: NN). This qualitative overview already displays clear differences in the resulting magnetic field: TL produces an artificial grid-like structure, due to an a priori higher field strength at the gridpoints. These artifacts are smoothed out using TC, since the eq.~\ref{eq:cubipol} can even provide an interpolated value that exceeds the values that are given at the grid points. As expected, NN produces a blocky image, where the edge length of each cell expresses the used spacing $d$ of the grid, since the same nearest neighbor value is used within a volume of $d^3$.
\\
More quantitative comparison of the different interpolation methods with respect to their properties are as follows:

First, the Fig.~\ref{fig:brms} shows the error of $B_{\rm rms}$ as a function of $l_{\min}/d$ for three different ratios of $l_{\min}/l_{\max}$. The interpolation routines TC (left panel) and TL (right panel) cause an error in the reproduction of the $B_{\rm rms}$ which is limited to below 10\%. It is the largest for small values of $l_{\min}/d$, increases with larger ratios $l_{\min}/l_{\max}$ and is generally larger for the TL method as compared to the TC approach. For fundamental reasons, NN does not yield any deviations except for the statistical uncertainties, so that it is not illustrated. In total, the impact of the interpolation routine is negligible for $l_{\rm min}/d\gg 1$ or $l_{\rm min}/l_{\rm max}\ll 1$.
Secondly, the standard deviation $\sigma_B$ of the flux distribution as a measure of $\text{div}(B)$ is shown in Fig.\ \ref{fig:fluxdist-staircase}. All methods perform worse on more coarse-grained grids (small $l_{\rm min}/d$) and for a small wave-mode range. In all cases, TC incurs a lower flux defect than both NN and TL, while NN outperforms TL without exception. Naturally, the gap between the methods shrinks as the flux defects get smaller; for changes in $l_{\rm min}/l_{\rm max}$, the relative difference remains constant.

\begin{figure}[htbp]
\centering
\includegraphics[width=0.8\linewidth]{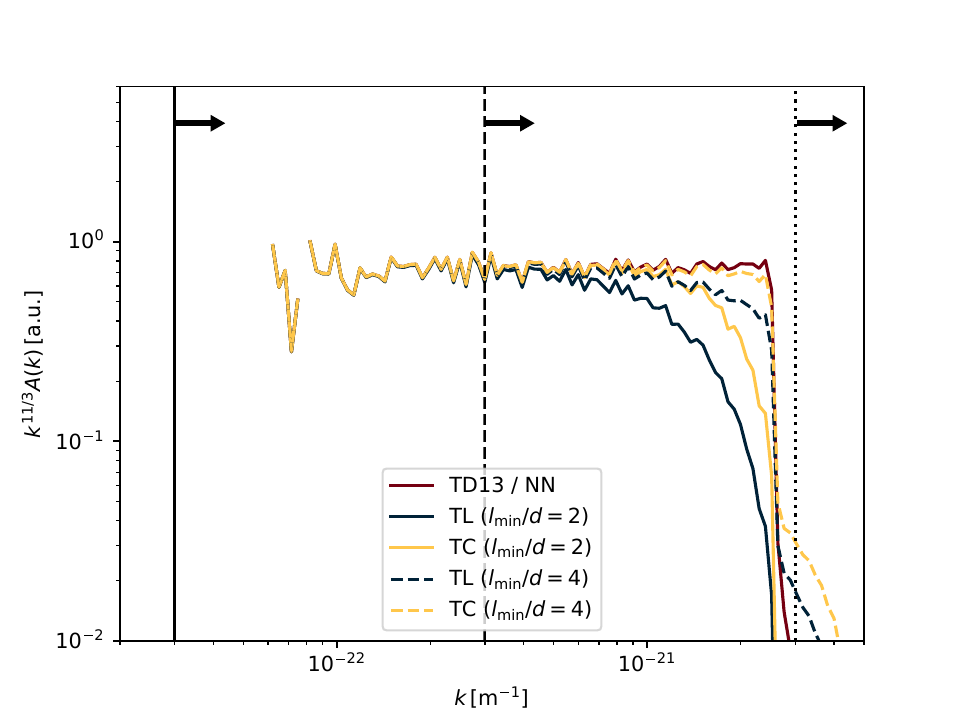}
\caption{Spectrum of the turbulent magnetic field in k-space in the case of $l_{\rm min}/l_{\rm max}= 0.017$ for two resolutions. \textbf{TD13}: a sampled TD13 field with 1200 wave-modes, \textbf{TL}: trilinear interpolation on the sampled TD13 field, \textbf{TC}: tricubic interpolation on the sampled TD13 field. The vertical lines mark the lower rigidity bound for the resonant scattering of particles with $\rho=1$ (\emph{solid line}), $\rho=0.1$ (\emph{dashed line}) and $\rho=0.01$ (\emph{dotted line}).}
\label{spectra}
\end{figure}

Finally, we analyze the resulting turbulence spectrum, as shown in Fig.~\ref{spectra}. 
Due to the discretization and windowing of the continuous TD13 field by the FFT, this routine provides a non-vanishing amplitude at $k>k_\mathrm{max}$ which, however, vanishes for an exact Fourier transform, i.e.\ an infinite number of transformation points. 
In addition, the limited number of wave-modes leads to an additional fluctuation resulting in a deviation from the desired amplitude $A(k)$. 
In this analysis, NN matches the behavior of the source field exactly, which is due to the fact that the source field is sampled only once, and the resulting grid is then used in both the direct FFT and as the basis for the interpolated FFTs. Since NN does not touch these values in any way, the resulting spectrum is exactly the same.

In contrast, TC and especially TL show a significant impact on spectral behavior, steepening the slope of the supposed power-law spectrum toward higher wave numbers. \Bjoern{Hence, at length scales smaller than the grid resolution, the spectral behavior is just a result of the interpolation method and a higher grid resolution is needed to account for the impact of small-scale turbulence, as shown in Fig.~\ref{spectra}.} 
\Bjoern{Otherwise, the} change of slope reduces the number of wave-modes the particles are able to scatter off. \Bjoern{But since the turbulence spectrum usually shows a descending slope, the diffusive transport is dominated by the large-scale wave-modes.} Although, for a rigidity $\rho\sim l_{\rm min}/l_{\rm max}$, i.e.\ close to the limit of the resonant scattering regime, the diffusive behavior is expect to change due to the steeper spectral slope by the interpolation routine, as analyzed in more detail in the following section. \Bjoern{Further, in the case of a dominant background field---which has not been investigated in this work---the spectral index of the turbulence spectrum is committed to the energy distribution of the test particles in the case of diffusive transport, so that a high grid resolution is needed to provide an accurate energy distribution.}  
\subsection{Propagation in the ballistic regime}
\begin{figure}[htbp]
\centering
\includegraphics[width=1.0\linewidth]{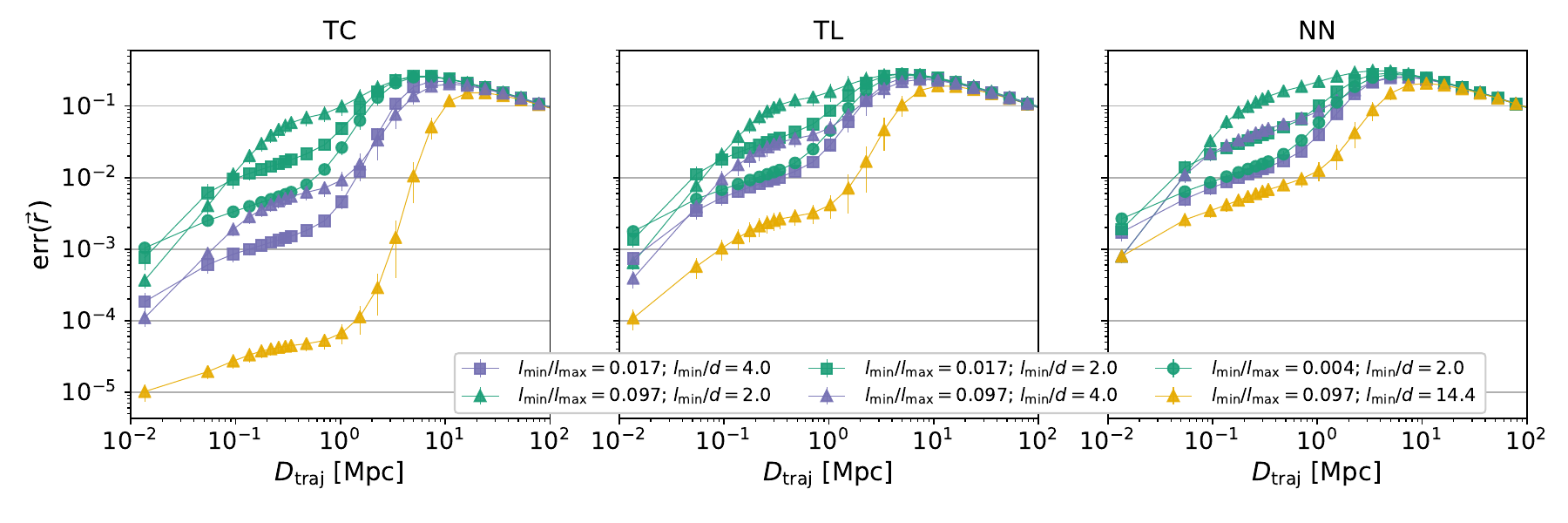}
\caption{Six different simulation cases are used to present the error in the particles' trajectory for different interpolation routines using particles with $\rho=0.1$. Each case uses a fixed combination of the ratios $l_\mathrm{min}/d$ and $l_\mathrm{min}/l_\mathrm{max}$.} 
\label{trajecComp01}
\end{figure}

\begin{figure}[htbp]
\centering
\includegraphics[width=1.0\linewidth]{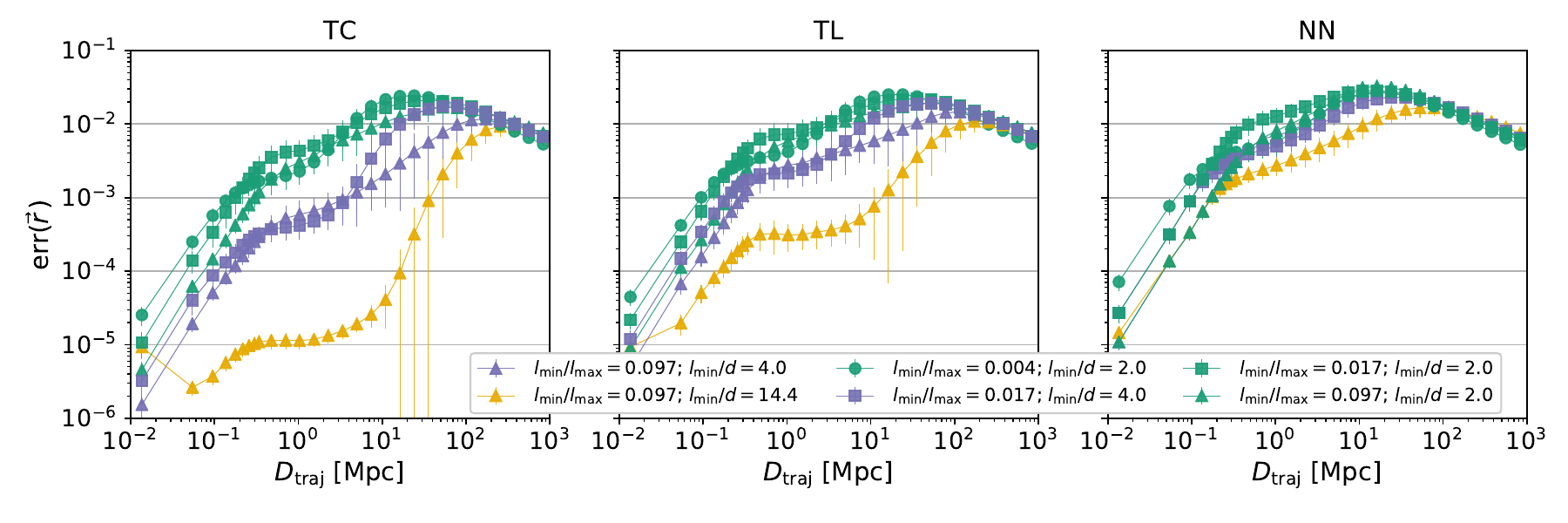}
\caption{Six different simulation cases are used to present the error in the particles' trajectory for different interpolation routines using particles with $\rho=0.01$. Each case uses a fixed combination of the ratios $l_\mathrm{min}/d$ and $l_\mathrm{min}/l_\mathrm{max}$.} 
\label{trajecComp001}
\end{figure}

Comparing the trajectories of 5000 isotropically emitted particles in 20 random configurations of the turbulent magnetic field, we determine the difference between the particles' positions $\vec{r}_{\rm TD13}$ and $\vec{r}$ using the analytical TD13 field and the grid based field with one of the interpolation routines, respectively. Dividing this difference by the particles' trajectory lengths yields the relative error as introduced by Eq.~\ref{relErrorTraj}. For all interpolation routines, Fig.\ \ref{trajecComp01} and \ref{trajecComp001} show that $\text{err}(\vec{r})$ is increasing up to a trajectory length where the particles are entering the regime of diffusive motion. In the case of $\rho=0.1$ this transition occurs at about $10\,\text{Mpc}$, whereas for $\rho=0.01$ it occurs at about ten times larger trajectory lengths. Here, the differences between the different interpolation methods and grid configurations vanish and the relative error decreases according to
\begin{equation}
    \frac{| \vec{r}_{\rm TD13} - \vec{r}|}{D_{\rm traj}}\propto D_{\rm traj}^{-1/2}\,,
\end{equation}
as expected from theory (\ref{diff_posDiff}). Hence, the relative error by the interpolation routine peaks with about $(10-20)\%$ for $\rho=0.1$ and $(1-2)\%$ for $\rho=0.01$ at about the transition between the ballistic and the diffusive regime. Hereby, the maximal value hardly depends on the used interpolation routine. In principle, the error predominantly depends on the grid resolution and hardly on the $l_{\rm min}/l_{\rm max}$ configuration. In contrast to the previous results on the magnetic field properties the NN method provides the largest errors on the particle trajectory with respect to the other two routines. Thus, the single trajectory comparison exposes that the use of a sophisticated interpolation algorithm as well as a high grid resolution minimizes the error on the particle's trajectory in the ballistic regime. But the maximal error at about the transition between the ballistic and the diffusive regime, is hardly effected by the interpolation routine or the grid configuration.
\subsection{Propagation in the diffusive regime}
\begin{figure}[htbp]
\centering
\includegraphics[width=1.0\linewidth]{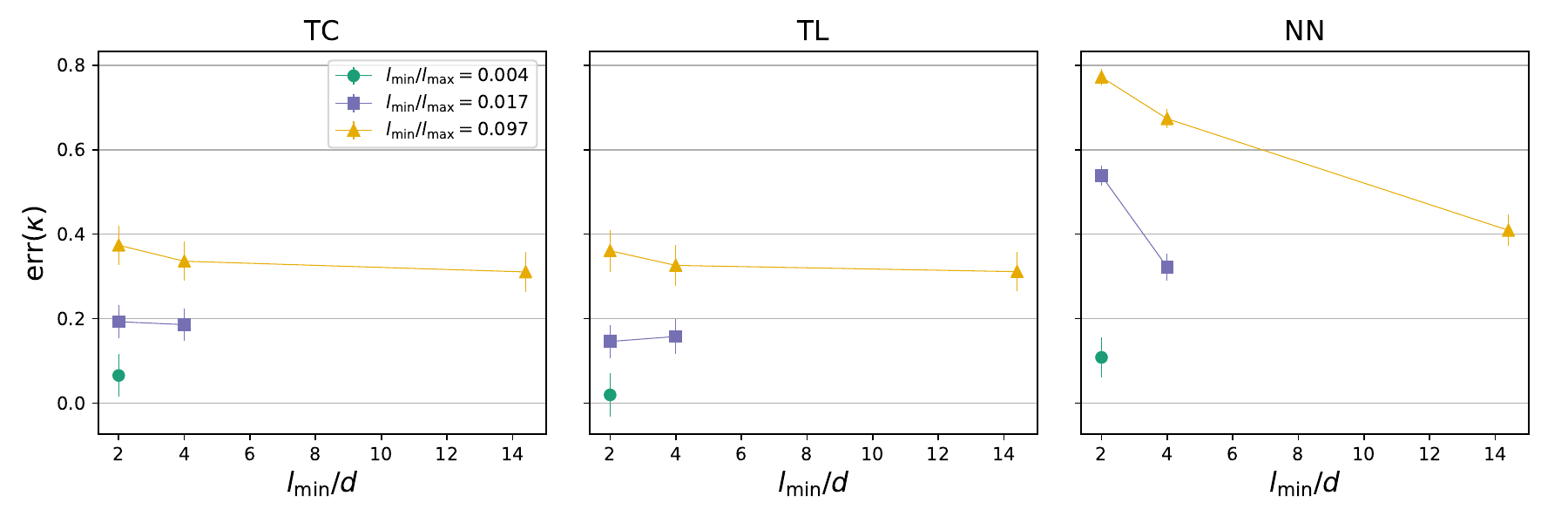} 
\caption{The relative error of the diffusion coefficients for different interpolation routines and field configurations using test particles with $\rho=0.01$.} 
\label{Diff_Cases001}
\end{figure}

\begin{figure}[htbp]
\centering
\includegraphics[width=1.0\linewidth]{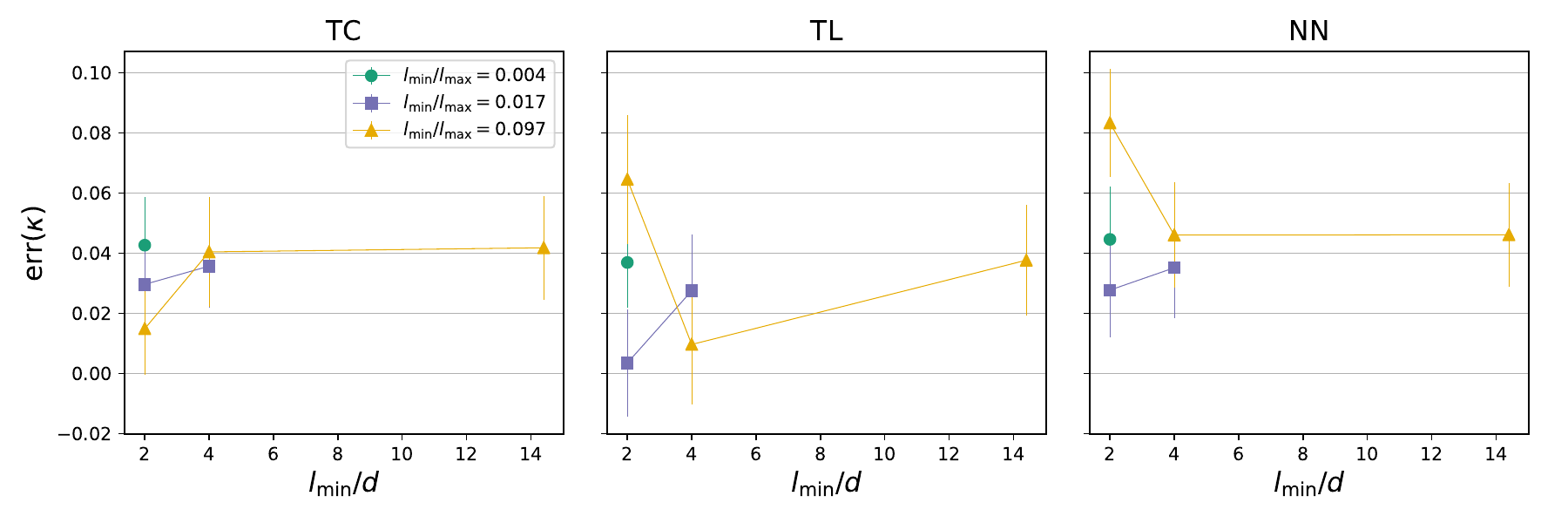} 
\caption{The relative error of the diffusion coefficients for different interpolation routines and field configurations using test particles with $\rho=0.1$.} 
\label{Diff_Cases01}
\end{figure}

In the diffusive regime, the single particle comparison has shown no significant differences dependent on the interpolation routines yielding $\text{err}(\vec{r})\lesssim 0.1$. To analyze the corresponding error of the diffusion coefficient, we apply only the last 5\% of the data of the mean distance $\left\langle (\Delta x_i)^2 \right \rangle$ of 5000 particles as discussed in Sect.~\ref{Sec:Method}. Using the TD13 field with 1200 wave-modes to provide the reference values of the diffusion coefficient, the relative error of the diffusion coefficient from interpolation methods is determined according to Eq.~\ref{relErrorDiffCoeff}. 
The Fig.~\ref{Diff_Cases001} and \ref{Diff_Cases01} show that the resulting error is hardly dependent on the interpolation method or the grid configuration, in particular in the resonant scattering regime, i.e. for all cases that satisfy Eq.~\ref{resonantScattering}. Here, the resulting diffusion coefficient $\kappa$ is in pretty good agreement with the expectations, as the accuracy of $\kappa_{\rm TD13}$ is only at the order of a few percentage. 
Further, the Fig.~\ref{Diff_Cases001} shows that for $\rho<l_{\rm min}/l_{\rm max}$ the error becomes quite huge, in particular for the NN interpolation method in the case of a low grid resolution.  Since the field line random walk dominates the particle transport in this rigidity regime, the error gets dominated by the accuracy of the interpolated magnetic field line on small scales. Therefore, TC and TL provide a significantly smaller $\text{err}(\kappa)$ than NN, although $\text{err}(\kappa)$ decreases significantly with increasing grid resolution for this interpolation method. For $\rho< l_{\rm min}/l_{\rm max}$ and $l_{\rm min}/d\leq 5$, the particle's Larmor radius is smaller than the grid resolution, i.e.\ $R_{\rm L}< d$, so that the small-scale change of the turbulent magnetic field cannot be resolved. Thus, the test particle essentially follows a magnetic field line that is smoothed in the case of TL or TC and unchanged in the case of NN. Note, that we do not account for the wave-modes of the dissipation range (at $k>k_{\rm max}$). Hence, in the case of a more realistic turbulence spectrum, we expect that the errors at $\rho< l_{\rm min}/l_{\rm max}$ slightly decreases.

\section{Conclusions}
Turbulent magnetic fields that provide the dominant magnetic field structure in a multitude of astrophysical environments are commonly generated by an Inverse Fast Fourier Transformation 
on a homogeneous, spatial grid. Due to this discretization, the magnetic field at an arbitrary spatial position requires the use of an interpolation routine. In this work, we determine the resulting error of three different interpolation routines (TL, TC and NN) with respect to the magnetic field properties and quantify its impact on the propagation of a test particle. 
The large-scale properties of the magnetic field are reproduced more accurately by the TC method than by the TL method, but even the NN method excels the TL routine. However, latter is not able to describe the small-scale properties of the turbulent magnetic field appropriately, which becomes important when field line random walk dominates the particle transport. Hence, the analysis of the particle propagation shows a quite diverse picture:
In the ballistic propagation regime (at $c\,t\lesssim \lambda$), the relative error by the interpolated field is in principle small, but increases in all cases up to a few percentage --- dependent on the particle's rigidity --- at $c\,t\sim \lambda$. Basically, TC excels TL, which excels NN, and an increasing grid resolution decreases the error. It needs to be taken into account, that the initial deviation from the expected spatial position is afterwards amplified by the turbulent magnetic field structure. Therefore, the error becomes the largest at the transition between the ballistic and the diffusive regime. In the diffusive regime (at $c\,t\gg \lambda$) for the case of resonant scattering, i.e.\ $1 \gtrsim \rho \gtrsim l_{\rm min}/l_{\rm max}$, the diffusive scattering off the wave-modes dominates the transport and the particle propagation is barely effected by the interpolation routine or the field configuration. However, when field line random walk dominates the transport at $\rho < l_{\rm min}/l_{\rm max}$, the interpolated field has a significant influence on the particle transport leading to an error of (20-80)\% of the corresponding diffusion coefficient. The NN method clearly provides the largest errors, but even in the case of a high grid resolution and the use of the TL method, we obtain a significantly different behavior of the particle transport. 

In total, the choice of the interpolation method clearly depends on the physical problem \Bjoern{that shall be addressed}: With respect to the particle propagation the TC method shows in principle the smallest uncertainties, but at the transition between ballistic and diffusive motion as well as for a dominating field line random walk, none of these methods is capable of keeping the uncertainty below 10\%. \Bjoern{So, even the most accurate interpolation routine yields at least a tiny deviation from the proper magnetic field seen by the test particle, which will be amplified exponentially when the quasi-perpendicular transport becomes diffusive. Thus, it is impossible to accurately follow the proper particle trajectory with a grid based magnetic field after some tens of gyrations. Here, the exact field, given by a continuous, grid-less approach like the TD13 method, is mandatory. However, if the average behavior of a particle ensemble is of interest, it is shown that the results do not depend significantly on the interpolation method as long as a sufficient grid resolution is used. Hence, }
for rigidities $\rho > l_{\rm min}/l_{\rm max}$, all interpolation routines can provide the diffusion coefficient with a precision of about 90\%. In the case of rigidities $\rho < l_{\rm min}/l_{\rm max}$ \Bjoern{a huge memory ($\gg 10\,\text{GB}$) is needed to provide the necessary grid resolution, so that the continuous, grid-less method is significantly less computationally expensive.}

We optimize the TD13 implementation, so that the necessary CPU time is reduced by more than an order of magnitude. Hence, in the case of 100 wave-modes, the performance of the optimized TD13 routine is equal to TL and only the use of NN is still slightly faster. Thus, the optimized TD13 routine with about 100 wave-modes and a length scale ratio $l_{\rm max}/l_{\rm min}$ of the turbulence at about the same order provides clear benefits compared to the grid-based approach: The turbulent magnetic field properties as well as the transport of particles can be described with a high accuracy and further, it has no technical difficulties to realize wave-modes on arbitrary scales. In contrast, the possible length scales of the resonant scattering regime is in the case of grid-based fields always limited by the available memory space. So it will becomes indispensable to use a grid-less field, if the particle propagation at very small rigidities $\rho\ll 0.01$ is considered within the resonant scattering regime.

Finally, we need to clarify, that the results of this work are based on the use of an isotropic turbulent magnetic field structure with a sharp cut-off beyond the inertial range of the spectrum. It is not expected that a more realistic turbulence spectrum yields significantly different results, as (i) the considered rigidities do not allow a resonant scattering with the wave-modes of the energy range (at $k<k_{\rm min}$) and (ii) only for $\rho < l_{\rm min}/l_{\rm max}$ do particles significantly interact with the wave-modes from the dissipation range. Further, it has been beyond the scope of this work to account for an underlying background field. However, we expect that some background fields amplify the uncertainty from the interpolation method significantly and suggest a detailed investigation using different background field configurations.

\acknowledgments
%
We are grateful to Andrej Dundovic for useful discussions as well as the comments of the unknown referee that helped to improve the original version of the paper.
L.S., A.F.\ and B.E.\ acknowledge financial support from the MERCUR project An-2017-0009.
%


\software{CRPropa3 \citep{CRPropa3}, NumPy \citep{vanDerWalt2011}, Matplotlib \citep{Hunter:2007}, SymPy \citep{sympy}, Pandas \citep{mckinney-proc-scipy-2010}.}

\bibliography{Bib}

\end{document}